# The Crisis in Astrophysics and Planetary Science: How Commercial Space and Program Design Principles will let us Escape

**Martin Elvis[1]**

*Harvard—Smithsonian Center for Astrophysics,*
*60 Garden St. Cambridge MA 02138, USA*
*E-mail:* `melvis@cfa.harvard.edu`

Astrophysics and planetary science are in crisis. The large missions we need for the next generation of observations cost too much to let us do more than one at a time. This spreads the science out onto a generational timescale, inhibiting progress in both fields. There are two escape paths. In the long run, but still well within our planning horizon, commercial space will bring mission costs down substantially allowing parallel missions at multiple wavelengths or to multiple destinations. In the short run, adopting prudent principles for designing a research program will let us maintain vitality in the field by retaining breadth at a modest cost in depth.

[1]Speaker

## 1. Introduction: the Crisis in Astrophysics and Planetary Science

The rapid growth of the cost of flagship missions in astrophysics (at about 10%/year adjusting for inflation [1]) and in planetary science (at about 15%/year, [2]) has created a crisis for both fields.

In astrophysics this crisis is obvious. Despite the clear message from virtually every object in the sky that they emit valuable signals across the entire electromagnetic spectrum, we will not be able to maintain our fleet of observatories matched in sensitivity in all bands. Instead cost is shrinking our capabilities to one waveband at a time. This will dramatically reduce the rate of progress in our understanding of the universe. Instead of waiting a year to follow-up a discovery in the optical with an observation in X-rays, for example, we must wait a decade.

It is at least as bad in planetary science. The 2011 US decadal survey for the field [3] listed three top priority destinations: Uranus, Europa and Mars. NASA could only afford to say "choose one". This automatically makes even this short list a generation-long program. Mars was chosen, with Europa following a decade later, most likely. Outer solar system science will suffer an almost career-length gap. Who then will be able to do that science if is re-started 20 years from now? Can we still say we are exploring the solar system when the pace is so slow? More realistically we are maintaining the capability to do so.

The prime example of a mission that has grown so expensive that it crowds out all other science in its field is the James Webb Space Telescope (JWST). This infrared (0.8 – 25 microns) cryogenic 6.5 meter diameter telescope will be an extraordinary machine. Compared with its predecessor, the 0.85 meter diameter Spitzer Space Telescope it will be orders of magnitude more capable in many parameters[2].

However at a cost to NASA of nearly $9 billion JWST is using nearly 20 years worth of the funding now available for large missions in the NASA astrophysics budget [4]. The wavelengths spanned by NASA's "Great Observatories" other than Spitzer, the Chandra X-ray Observatory and Hubble, or by ESA's Herschel telescope: i.e. the ultraviolet, X-ray, and far-infrared bands[3] must wait until about 2030, at the earliest, to be replaced. By that time Chandra, for example, will be 30 years old. Even if it is still operating it will be unable to match JWST in sensitivity.

---

[2] Note that JWST is often talked about as the replacement for the Hubble Space Telescope (HST), but they overlap only in the 0.8 – 1.6 micron band, which is a small section of either telescope's bandpass. It is an *intellectual* successor to HST in that it takes the study of galaxies to higher redshifts by moving to longer wavelengths, but in all of the other science areas where JWST and HST are highly valued, this argument does not apply.

[3] All of which must be observed from space.

This lack of a matched set of great observatories going forward is bad, even for JWST science. A set piece of the JWST observing program will be the deep fields. One of their main goals is to find the first galaxies where the first stars in the universe are forming. JWST may well detect these at redshifts of 10 or 20 (when the universe was only 170 – 450 Myr old). But how will anyone know if the infrared blobs they see are powered by the first stars rather than the first black holes? These black holes must be rapidly growing then in order to become the supermassive black holes we already see by redshift 6 (at an age of 890 Myr, compared with today's 13.7 Gyr age). A comparably deep X-ray image would decide this at once, but that must wait a decade or two with the program we have now, leaving this prime JWST science in limbo.

Growing costs for missions have trapped us into a serial approach to covering the electromagnetic spectrum rather than the parallel approach we have enjoyed for about four decades[4]. Is there a way out?

The way to escape this problem must be to bring down the costs. Below I discuss two ways out of the trap. The cost discipline provided by profit-making enterprises in space, commercial space, will force down prices. Significant changes are happening well within our current planning horizon. Independently, we can make our programs stronger by deliberately choosing a set of prudent science program design principles.

**2. Commercial Space will reduce mission costs**

For decades the inflation-adjusted cost of getting to low Earth orbit (LEO) was about $10 k/kg. That meant that there was little new for scientists to think about when it came to the capabilities we had in space. This is now changing fast.

**2.1 Launch Cost per kilogram**

Already SpaceX advertises rates that are $3 k/kg [5]. With re-useable first stages, which carry 9 out of 10 of the expensive engines on a Falcon 9, this cost is likely to drop another factor of 2. With Falcon 9 Heavy the advertised cost is about $1.6 k/kg to LEO, without first stage reuse. So it is now reasonable, even conservative, to plan for post-2025 missions using a cost to LEO that is about 1/5 the historical cost. As launch is of order 25% of the cost of a mission, this is a direct cut of 20% to the overall mission budget. This is a useful, but not revolutionary, saving. Even free launches would not allow another mission of comparable scale to be developed in parallel.

**2.2 Relaxed Mass Constraints on Spacecraft and Science Payload Design**

But cheaper mass-to-orbit will enable a fresh approach to spacecraft and science payload design. In spacecraft design keeping mass to a minimum is the overriding goal.

---

[4] Beginning with the International Ultraviolet Explorer (IUE, launched 1978, the Einstein Observatory (X-rays, launched 1978) and the InfraRed Astronomical Satellite (IRAS, launched 1983).

The inexorable rocket equation makes a linear increment in payload mass cost an exponential growth in total mass. Hence every 100 grams matters. This drives spacecraft design to extreme mass-savings that produces a rigorous design and complex test process. Each spacecraft must be designed precisely for the mission at hand. This is expensive. If mass constraints are eased a less costly process can follow.

There are studies [6] that suggest relaxing mass constraints slightly (e.g. 50% mass growth) could cut spacecraft costs by large factors (of order 3). As the spacecraft is roughly half the cost of an astrophysics mission this begins to make developing a second mission in parallel plausible. Margin to relax the mass constraint on the missions exists. With a payload mass to LEO of 22.8 mt [5] the Falcon 9 has excess capability over that needed to launch any of the recent Discovery or Explorer missions.

Even the scientific payload could benefit from a mass constraint relaxation. While the optics, sensors and pre-amp electronics must always be state-of-the-art, or the mission is not worth flying, the other payload systems offer major cost savings in a less mass-constrained environment [1].

Studying this idea afresh with today's technologies is likely to pay-off greatly for NASA.

**2.3 Low Cost On-orbit Science Payload Testing in LEO**

Commercial space is already helping science missions by providing a low cost means to test systems to the highest technical readiness level (TRL-9) aboard cubesats. Larger systems, including science payloads, may become testable at modest cost in the same way once commercial crew begins and, along with that, space tourism. The 14 $m^3$ trunk of the Space-X Dragon [5] is rarely used at present, but provides power, telemetry, and some limited pointing capabilities. A two week test mission is feasible, although the instrument will then typically burn-up on re-entry. The flight rate would grow further if commercial space stations, like the Bigelow Aerospace 330, were to become successful.

Post-2020 there may well then be frequent opportunities to use these flights for TRL-9 testing of large instruments as secondary payloads on crew or cargo missions. Compared with a few minute rocket flight, whose payloads also typically do not survive landing, the Dragon trunk offers greatly enhanced capabilities. In two weeks significant science could be achieved as well as instrument testing. Earlier adoption of advanced technologies should result, and lowered risk may enable even the science payloads to find cost savings.

It would be natural to fund this program as an extension of the NASA sub-orbital program, now used for rocket and balloon flights.

**2.4 Affordable On-orbit Servicing in LEO**

On-orbit servicing in LEO has an outstanding proven scientific track record with Hubble. But the cost in the Shuttle era was excessive, and later missions deliberately avoided the possibility to keep run-out costs down [1]. With the advent of commercial crew on-orbit servicing in LEO can come back into the running. Being able to replace failed components allows higher risk profiles to be contemplated, so that instrument failure becomes merely inconvenient, not mission-ending. Lower costs and more cutting-edge instruments should then be enabled. For now, at least, this option might re-orient missions in the planning stages toward LEO and away from Sun-Earth-L2.

**2.5 Commercial Lunar Landers**

Within five years there are likely to be several non-NASA lunar lander missions. In particular the landers developed for the Google Lunar-X-prize are planned to cost of order $50 M [7]. Multiple cheap landers could visit well-selected lunar sites based on the past decade of multi-technique imaging of the Moon [8], to the great benefit of planetary science.

Astrophysics could use this newly affordable capability, not only to emplace (1) the long-discussed far-side low frequency radio telescope, but also (2) to establish a stable baseline sub-millimeter VLBI antenna to supplement the Event Horizon Telescope, enabling the imaging of dozens of supermassive black holes, and (3) high angular resolution far-infrared interferometric arrays in the multi-kilometer scale 'cold traps' at ~30 K, in the permanently dark craters at the lunar poles [9].

**2.6 The Prospect in 2025**

All of this adds up to radically altered landscape as we contemplate what scientific space missions to put forward for the late 2020s and beyond (Table 1). It would not be wise to ignore these developments.

Table 1: Commercial Space activities effects on Astronomy and Planetary Science to 2025 [1]

| COMMERCIAL SPACE ACTIVITY | EFFECT ON COST | VALUE TO ASTRONOMY, PLANETARY SCIENCE |
|---|---|---|
| Lower cost launch (≥2X) | Cheaper Spacecraft. (factor 2-3?, "off-the-rack") | "2 for 1" missions, both astronomy and planetary science |
|  | Cheaper science payloads (factor 2?) |  |
| Passenger flights to LEO | Cheap High TRL tests | Cutting edge instruments |
|  | Cost-effective servicing in LEO | Higher risk sub-systems in LEO |
| Private Space Stations | More passenger traffic | Indirect via more passenger flights |
| Private Lunar Landers | Cheaper access to lunar surface | Many expeditions to well-chosen sites; radio, sub-mm, far-IR telescopes |

**2.7 How to get there**

Achieving low cost missions involves a space agency using a novel cost model. Adopting a new cost model is a big and risky step, especially when a flagship mission is at stake. NASA and the other space agencies will rightly demand convincing evidence that this new approach is both truly cheaper and will deliver reliable spacecraft. The probable path is that commercial satellites will first demonstrate that the approach works, and the industry will have suitable spacecraft ready for purchase. Space agencies will likely then try out the approach on medium-class missions first, before committing to building flagship missions this way. This approach would delay the gains for astronomy and planetary science, but would make them more certain in the longer run.

**3. Prudent Program Design Principles**

But even if we can make a mission cheaper, how do we avoid the strong tendency to go for the One Big Mission? Prudence suggests that we adopt some clear principles for the design of a science program.

Without guiding principles, for any scientist, the temptation will be strong to use any savings on launch and spacecraft to enhance the science payload: "Payloads grow to fill the budget available", you might say. This is the path of least resistance. To restrain the scientists from spending all the gains on more payload rather than on a second flagship will need discipline.

This may be a discipline mandated by the space agencies, or it may be a self-imposed discipline based on a community consensus that more missions are better than one. Of course, each one still needs to be a big step over its predecessor. A combination of both approaches may be required. Without community buy-in agency rules may not affect the decadal priorities; without imposed rules the community will not be challenged to become more creative.

There are several arguments that a large number of missions are better than one giant mission. These arguments are extensions of well-accepted mission design criteria applied to an entire science program. With these principles in place they can be flowed down to particular missions and sets of missions.

*3.1 Single-point failure*

This is a concept used in designing missions to ensure that there is no one component, sub-system or system in the mission that could be mission-ending if it fails. The same consideration applies to the entire science program. If we "put all our eggs in one basket" by going all out for a single giant mission, then we expose the science program to single point failure risk. Had on-orbit servicing not been possible to correct the spherical aberration in its optics, Hubble would have been, if not a failure, not a great success. A program with a built-in single point failure is not robust.

*3.2 Scientific Program Requirements*

The idea of developing a set of science requirements, and then flowing them down into mission requirements, is standard in designing a mission; it should be no less so in designing an entire science program.

The need to have matched spectrum-spanning capabilities operating in parallel for $21^{st}$ century astrophysics is a clear example [1]. A program that does not enable this multi-wavelength astrophysics is not fulfilling its basic mandate.

A single large mission comes with an opportunity cost. What missions are not being done because this one was chosen? Without this tensioning between options, there is always a tendency to say yes when a large mission is considered. Making the opportunity cost of every mission explicit needs to be a requirement when it is considered as part of an overall science program.

A part of the science requirements for a program is to be always trying something new. Whenever the cost of a single project becomes too large, the risk-taking inherent in any new approach tends to stifle innovation. This is not a problem unique to space missions. The movie business has the same issue. Filmmaker Patricia Rozema says of moviemaking: *"It's an inherently conservative business because it's so expensive. And if you're not repeating something that's already a success then people are nervous."* [10]. To counter this conservatism a space program needs to promote low cost "indies" through, for example, more vigorous Explorer and Discovery programs [11], and preferably by making even ambitious missions affordable.

*3.3 Single viewpoint failure*

Equally important is the intellectual vitality of the field. If all astronomers use just one telescope, the breadth of their scientific imagination will be limited, and their results less challenged by independent data. Scientists dependent on being granted time (and funding) on a single telescope will feel a pressure to go along with current fashions in what questions and approaches are important. Time assignment committees tend to be made up of those who were successful in earlier rounds, reinforcing those topics as appropriate for the field.

These effects, though unintentional, tend to restrict creative new approaches. Exoplanets are an example. Both Hubble and Spitzer now have large programs devoted to exo-planet science. It is hard to believe that speculative searches for exo-planets would have been pioneered by these telescopes, however, given the intense competition for their observing time, and their existing user communities. Instead the ability of small teams of astronomers to take a different path with lower cost telescopes led to an explosion of new science, one that was utterly unanticipated by the large majority of astronomers [12].

### 3.4 Implementation

Following these principles will lead to a program that deliberately avoids pushing a single big mission.

### 3.4.1 Balancing Ambitious, Achievable and Affordable

The design of a new mission is always a balance between the unlimited demands of scientists with the practical capabilities of the latest technologies. The art is in balancing these demands. Each mission needs to be "ambitious, achievable and affordable": the goal must be a major step beyond what has already been done. Technical readiness needs to be high before committing to large missions. And the cost must not squeeze out all other comparble missions.

An advantage of not having one big mission is that, with a number of major missions under consideration, it is easier to delay – or even cancel - one that has not demonstrated a sufficient TRL for key systems. Support for technology development at a substantial level then needs to be a part of the program. Some 'not to exceed' fraction of the mission cost may be appropriate, with continued funding dependent on successful progress reviews.

It is all too easy to think of a science goal that needs a telescope far bigger than we currently have. The High Definition Space Telescope (HDST)[5], for example, has the great and inspiring goal of finding the signatures of life on exo-Earths [13]. But, in my judgement [14], it makes a poor case that it can do so, and at an excessive cost. It really needs a more powerful telescope, and perhaps an utterly different technological approach. For me, the science is premature. (I wish it were otherwise because I'd love to know the answer.)

In general, if the science goal costs a generation's worth of funding, then you have made the wrong choice. Realizing when a new technology can provide order-of-magnitude advances at a reasonable cost is where the skill comes into mission design. There are presently several windows of the spectrum where promising giant leaps may be made at relatively modest cost.

### 3.4.2 Community Challenges

The Decadal review process is the means by which the community produces a plan for the coming decade. Steven Squyres [15] has noted that the two most important elements for a successful decadal survey are (1) the statement of task (SOT), and (2) the decision rules. Incorporating the program design principles discussed here into the decadal SOT and into the decision rules would create a strong framework for that survey. (The principles are not limited to space-based programs, but can apply to any science program.)

---

[5] HDST is a candidate for the Large Ultraviolet, Optical, Infrared Surveyor (LUVOIR) mission.

Sociologically it is harder to push for multiple diverse missions than for one giant mission. A giant mission proposal tends to accrete a large number of scientists, and this unified large community then speaks louder than several smaller ones, even if these smaller ones are more numerous in aggregate. Packaging a suite of spacecraft into a single program could have the same effect as pushing a single giant spacecraft.

Hubble, Chandra and Spitzer were successfully sold in the mid-1980s as such a package, the "Great Observatories". This was at the time a gamble. We knew too little to be sure that the three were well matched in sensitivity. Somewhat by luck they turned out to be excellently matched and were, and continue to be, wildly successful.

### 3.4.3 Agency and Government Buy-in

Politically it is also hard to get backing for a list of excellent missions both to the space agencies and to the governments that will pay for them. Saying these are the "two best" missions has intrinsically less force than saying this is "the best" mission. A "best program" appoach, mirroring that of the Great Obsevatories, and calling for a team of "Greater Observatories"could be the answer.

The astronomy and planetary science communities should actively consider and debate these issues.

## 4. Why not ask for more?

Why should we not ask for more? Oliver Twist tried this, wanting more gruel in the orphanage [16]. He was not successful. Science in general is not an easy sell once it exceeds ome cost threshold where it appears to be eating into other government supported activities. A "useless" science, even one as photogenic as astrophysics, is particularly hard to push through. But this is no reason not to try. The Great Observatories program did, after all, lead to a long-term growth in the NASA astrophysics budget. Some increase is surely possible.

However, asking for $9 B for a single telescope is already tough. To complete that within a decade, so that a second can fly while the first is still operational, we need to double the large mission budget for NASA astrophysics. If we want to build two flagship telescopes within a decade in order to span the spectrum in a 20-year program then we must *quadruple* the same budget line. (And that still omits the pressing new science of exoplanets and gravitational waves.) Following this path we soon get into a realm of unbelievable growth. Cost reduction is the only way.

Or is an alternative to go global and unite all the world's scientific space programs into a single co-ordinated whole? The history of joint NASA-ESA ventures is not encouraging. After a series of unfortunate events in 2010-2011 due to out-of-synch budgetary procedures (and, privately, some suggestions of unfair play) the two agencies decided that they would limit any other agency to a minority share in any of their missions. This limit is informal, and was implemented to protect each mission from cancellation due to the partner agency failing to come through with their contribution. Making joint-mission agreements with other partners is likely to be at least as difficult.

Each agency could instead agree to build one of the "Greater Observatories" with observation and data sharing agreements with the other agencies. Mission scope, launch dates, lifetimes and orbits would all need co-ordinating if the fully synergy of the missions is to be realized. This is not impossible, but it is highly challenging.

Even if successful, though, this global alliance is only a one-off, factor 2-3 increase in budget over NASA alone. It is not an open-ended path forward.

The fundamental mis-match between the cost growth rate of astrophysics and planetary science missions (10% - 15% [1,2]) and that of the underlying economy (1% - 3% [17]) remains large. Eventually we will hit this funding wall. Best to bite the bullet and take advantage of the new developments in commercial space while they are fresh.

## 5. Conclusions

The cost of scientific space missions is growing far faster than the underlying economy that supports them. This has created a "funding wall" crisis for both astrophysics and planetary science: we can no longer afford the suite of missions that the science requires. Both fields are in danger of intellectual decay. A larger budget at the scale needed is, at best, implausible. Banding together globally without changing the underlying cost structure is hard and will only make a temporary dent in the problem.

There are two ways to escape this crisis:
*First:* take advantage of the rapid, near-term, changes in commercial space activities to cut the costs of launches, spacecraft and science payloads. Launch costs are already dropping to about 1/5 of historical levels. Using this cheaper mass to orbit, spacecraft could cut costs by factors of a few, but how to achieve this improvement needs to be revisited, as do the potential cost savings on science payloads. Commercial crew flights open up the possibility of cost-effective TRL-9 testing of large instruments and on-orbit servicing in LEO. Both of these allow for the use of more cutting-edge instruments, and cost-effective servicing can allow higher risk mission profiles. The effects of these developments combine to allow multiple flagship missions to be built in parallel.

*Second:* adopt science program design principles that mirror those of mission design principles. Key among these are: (1) no single point failure; (2) scientific program requirements; (3) no single *viewpoint* failure. These principles require rejecting the tendency to select one big mission, and instead embracing more, smaller – but still order-of-magnitude advancing – missions.

With these approaches the space agencies will be well-positioned to ride the cost-reducing, capability-enhancing wave of expanding commercial space activities.

I thank the Aspen Center for Physics funded by NSF grant # 1066293 for their hospitality when this paper was begun.